\def\rfr#1{eq. (\ref{#1})}

\def\virg#1{``#1''}

\def\eqi{\begin{equation}}
\def\eqf{\end{equation}}
\def\eqia{\begin{eqnarray}}
\def\eqfa{\end{eqnarray}}
\def\rp#1#2{{#1\over#2}} \def\lb#1{\label{#1}}

\documentclass[11pt]{article}
\usepackage{amsmath,amsthm,amscd,amssymb}
\usepackage{latexsym}
\usepackage{graphicx,epsfig}
\usepackage{lscape,rotating}
\begin{document}

\noindent{\bf \LARGE{The perihelion precession of Saturn, planet X/Nemesis and MOND}}
\\
\\
\\
{Lorenzo Iorio}\\
{\it INFN-Sezione di Pisa\\
Viale Unit$\grave{a}$ di Italia 68, 70125\\Bari (BA), Italy
\\tel. 0039 328 6128815
\\e-mail: lorenzo.iorio@libero.it}

\begin{abstract}
We show that the anomalous retrograde perihelion precession of Saturn $\Delta\dot\varpi$, recently estimated by different teams of astronomers by processing ranging data from the Cassini spacecraft and amounting to some milliarcseconds per century, can be explained in terms of a localized, distant body X, not yet directly discovered. From the determination of its tidal parameter $\mathcal{K}\equiv GM_{\rm X}/r_{\rm X}^3$ as a function of its ecliptic longitude $\lambda_{\rm X}$ and latitude $\beta_{\rm X}$, we calculate the distance at which X may exist for different values of its mass, ranging from the size of Mars to that of the Sun. The minimum distance would occur for X located perpendicularly to the ecliptic, while the maximum distance is for X lying in the ecliptic. We find for rock-ice planets of the size of Mars and the Earth that they would be at about 80-150 au, respectively, while a Jupiter-sized gaseous giant would be at approximately 1 kau. A typical brown dwarf would be located at about 4 kau, while an object with the mass of the Sun would be at approximately 10 kau, so that it could not  be Nemesis for which a solar mass and a heliocentric distance of about 88 kau are predicted.
If X was directed towards a specific direction, i.e. that of the Galactic Center, it would mimick the action of  a recently proposed form of the External Field Effect (EFE) in the framework of the MOdified Newtonian Dynamics (MOND).
\end{abstract}

{\it Key words}: Experimental studies of gravity \*\ Experimental tests of gravitational theories  \\
{\it PACS}: 04.80.-y,04.80.Cc

\section{Introduction}
 Anderson et al. \cite{And09} recently examined some still unexplained anomalies connected with astrometric data in the solar system.
They are the flyby anomaly \cite{And08}, the Pioneer anomaly \cite{And98}, the secular change of the Astronomical Unit \cite{Kra04} and the increase in the eccentricity of the Moon's orbit \cite{Wil08}.
In fact, there is the possibility that also a fifth anomaly does actually exist: the anomalous perihelion precession of Saturn \cite{Ior09}.

The corrections $\Delta\dot\varpi$ to the standard Newtonian/Einsteinian secular precession of the longitude of the perihelion\footnote{$\varpi=\omega+\Omega$, where $\omega$ is the argument of perihelion and $\Omega$ is the longitude of the ascending node, is a \virg{dogleg} angle \cite{Roy05}.} $\varpi$ of Saturn, estimated with the latest versions of the EPM \cite{Pit09} and INPOP \cite{INPOP}
ephemerides by including some years of continuous radiometric ranging data to Cassini in addition to data of several types spanning the last century, are\footnote{The formal, statistical error in the Pitjeva's result is 0.7 mas cy$^{-1}$; Pitjeva (E.V. Pitjeva, private communication, 2008) warns that the realistic uncertainty may be up to 10 times larger. Anyway, she released the figure quoted in \rfr{pitperi}, also cited in Ref. \cite{Fie09}.} \cite{Pit08,Fie09}
\begin{eqnarray}
  \Delta\dot\varpi_{\rm Pit} &=& -6\pm 2\ {\rm mas\ cy^{-1}}, \lb{pitperi}\\
  \Delta\dot\varpi_{\rm Fie} &=& -10\pm 8\ {\rm mas\ cy^{-1}}\lb{fieperi};
\end{eqnarray}
both are non-zero at a statistically significant level ($3\sigma$ and $1.2\sigma$, respectively) and they are compatible each other since  their difference is equal to $4\pm 10$ mas cy$^{-1}$. At the moment\footnote{Plans to improve the ephemerides of Saturn with VLBA observations to Cassini exist \cite{VLBA}.}, no corrections $\Delta\dot\varpi$ estimated with the DE ephemerides by  NASA JPL are available.
Iorio in Ref. \cite{Ior09} unsuccessfully examined  several possible dynamical explanations  in terms of both mundane, standard Newtonian/relativistic gravitational physics and of modified models of gravity. Anyway, further analyses of extended data sets from Cassini with different dynamical force models are required to firmly establish the existence of the anomalous perihelion precession of Saturn as a genuine physical effect.

Here we will show that the existence of a localized distant  body (planet X/Nemesis), modeled in neither EPM nor INPOP ephemerides, is a good candidate to explain a secular perihelion precession of Saturn having the characteristics of \rfr{pitperi}-\rfr{fieperi}: indeed, contrary to a massive ring usually adopted to model the action of the minor asteroids and of the Trans Neptunian Objects (TNOs), it yields a retrograde secular perihelion precessions and the constraints on its distance for different postulated values of its mass are consistent with several theoretical predictions put forth to accommodate some features of the Edgeworth-Kuiper belt \cite{Lyk}. Concerning Nemesis, it would be an undiscovered stellar companion of the Sun  which, moving along a highly elliptical orbit\footnote{It should have semimajor axis $a=88$ kau and eccentricity $e>0.9$ \cite{Nem1}.}, would periodically disturb the Oort cloud being responsible of the  periodicity of about 26 Myr in extinction rates on the Earth over the last 250 Myr \cite{Nem1,Nem2}; the Nemesis hypothesis has also been used to explain the measurements of the ages of 155 lunar spherules from the Apollo 14 site \cite{Mul02}. See e.g. Ref. \cite{IorNem} for further details. Interestingly, such a proposed explanation of the anomalous perihelion precession of Saturn in terms of pointlike dark matter is, to a certain extent, to be considered as degenerate since also the MOdified Newtonian Dynamics (MOND) \cite{Mil83} predicts certain subtle effects in the planetary region of the solar system that may mimic the action of a distant mass\footnote{It should not be confused with the supermassive black hole in Sgr A$^{\ast}$ whose action on the solar system's planets is, as we will see, at present undetectable.} located in the direction of the Galactic Center (GC) \cite{Mil09}.
\section{The action of a distant body and of MOND's External Field Effect on the perihelion of a planet}
A hypothetical, still undiscovered  body X, located at heliocentric distance $r_{\rm X}\gg r$ along a direction ${\hat{n}}_{\rm X}$, where $r$ is the distance of a generic known planet P of the solar system, would impart on it a perturbing acceleration $\vec{A}_{\rm X}$ consisting of an \virg{elastic} Hooke-like term plus a term directed along $\hat{n}_{\rm X}$ \cite{IorNem}
\eqi \vec{A}_{\rm X} \approx -\mathcal{K}\vec{r} + 3\mathcal{K}\left(\vec{r}\cdot\hat{n}_{\rm X}\right)\hat{n}_{\rm X},\lb{acce}\eqf
where
\eqi\mathcal{K}\equiv \rp{GM_{\rm X}}{r^3_{\rm X}}\eqf
is the so-called tidal parameter of X.
Note that the acceleration of \rfr{acce} derives from the following quadrupolar potential \cite{Hog91}
\eqi U_{\rm X}\approx\rp{\mathcal{K}}{2}[r^2 -3(\vec{r}\cdot\hat{n}_{\rm X})^2].\lb{pot}\eqf

Iorio in Ref. \cite{IorNem} worked out the orbital effects of \rfr{acce} on the longitude of perihelion $\varpi$ of a planet P by means of the standard Gauss perturbing approach with the assumption that $\vec{r}_{\rm X}$ can be considered constant during an orbital revolution of P.
More specifically, the Gauss equation for the variation of the longitude of perihelion $\varpi$ of a planet under the action a small perturbing acceleration $\vec{A}$, whatever its physical origin may be, is \cite{Ber03}
\eqi\dot\varpi = \rp{\eta}{nae}\left[-A_r\cos f+A_{\tau}\left(1+\rp{r}{p}\right)\sin f\right]+2\sin^2\left(\rp{I}{2}\right)\dot\Omega,\eqf
where $A_r,A_{\tau}$ are the radial and transverse components of $\vec{A}$, respectively, $a,e,I$ are the semimajor axis, eccentricity and inclination, respectively, $\eta=\sqrt{1-e^2}$, $n=\sqrt{GM/a^3}$ is the unperturbed Keplerian mean motion, $p=a(1-e^2)$ is the semi-latus rectum, and $f$ is the true anomaly counted positive anticlockwise from the perihelion. The Gauss variation equation for the node $\Omega$ is \cite{Ber03}
\eqi\dot\Omega = \rp{1}{na\sin I\sqrt{1-e^2}}A_{\nu}\left(\rp{r}{a}\right)\sin (\omega+f),\eqf where $A_{\nu}$ is the normal component of $\vec{A}$.
It must be recalled that, in order to make meaningful comparisons with the estimated corrections $\Delta\dot\varpi$ to the standard perihelion rates, they have been obtained by processing the planetary data in the standard ICRF frame which is a frame with the origin in the (known) solar system's barycenter and having the mean ecliptic at J2000 epoch as reference plane with the $x$ axis directed towards the Vernal point \cite{Joh99}. This is particularly important when there is some physical feature, like a static body in a given direction as in our case, which breaks the spatial symmetry: assuming that $\hat{n}_{\rm X}$, which is actually a-priori unknown, coincides with one of the frame's axes just to simplify the calculation is, in principle, incorrect. The analytical calculations for a non-privileged direction of $\hat{n}_{\rm X}$ are very cumbersome\footnote{The software MATHEMATICA has been used.}; at the end, one is left with an expression of the kind
\eqi\left\langle\dot\varpi\right\rangle_{\rm X}=\mathcal{K}F(a,e,I,\Omega,\omega; \lambda_{\rm X},\beta_{\rm X}),\lb{mega}\eqf where $F$ is of the form
 \eqi F=\sum_i G_i(a,e)T_i(I,\Omega,\omega;\lambda_{\rm X},\beta_{\rm X})\eqf in which $G_i$ are complicated functions of the semimajor axis $a$ and the eccentricity $e$ of P, while $T_i$ are trigonometric functions of the inclination $I$, the longitude of the ascending node $\Omega$ and the argument of perihelion $\omega$ of the planet P perturbed by X, whose ecliptic longitude and latitude are $\lambda_{\rm X}$ and $\beta_{\rm X}$. Releasing such analytic expressions would be, actually, extremely space-consuming and of little help: they have to be numerically computed for given values of $a,e,I,\Omega,\omega$.
 As a result, by comparing \rfr{mega} to the estimated correction $\Delta\dot\varpi$ for a given planet P like Saturn, one has the tidal parameter $\mathcal{K}$ of X as a function of $\lambda_{\rm X},\beta_{\rm X}$.

At this point, it is interesting to note that the form assumed by the External Field Effect\footnote{It is one of the non-linear features of MOND according to which the internal dynamics of small system $s$, like the solar system, does depend on the external field of a larger system $S$, like the Milky Way, in which $s$ is embedded.} (EFE) in the planetary regions of the solar system in the recent study by  Milgrom \cite{Mil09}  has exactly the same functional dependence of \rfr{acce}, provided that
\eqi \mathcal{K}\rightarrow -\rp{q}{2}\left(\rp{A_0}{r_t}\right),\ r_t=\sqrt{\rp{GM_{\odot}}{A_0}}=6.833\ {\rm kau},\eqf
where \cite{Bege} $A_0=1.27\times 10^{-10}$ m s$^{-2}$ is the characteristic acceleration scale of MOND, and
\eqi \hat{n}_{\rm X}\rightarrow -\hat{x}.\eqf Indeed, Milgrom in Ref. \cite{Mil09} uses a frame with one coordinate axis directed towards\footnote{The Galactic external field is, indeed, the source of the centripetal acceleration $A_c\approx A_0$ of the Sun during its motion of revolution around GC.} GC; as we will see below, it is approximately the opposite of the $x$ axis of ICRF\footnote{It is the $z$ axis in the frame used by Milgrom in Ref. \cite{Mil09}.}. Although such an effect would manifest itself in the strong-field regime existing in the planetary regions of the solar system,  the functional form of $q$ depends on the form of the MONDian interpolating function\footnote{$X$ is the ratio of the total gravitational acceleration felt by a body to $A_0$.} $\mu(X)$ in the transition region in which $X\sim 1$, i.e. approximately at $r_t$.

Let us start to examine just the case of a dark object placed in the same direction of GC.
The right ascension $\alpha$ and declination $\delta$ of GC, assumed coincident with Sgr A$^{\ast}$, are \cite{Rei04}
\begin{eqnarray}
  \alpha_{\rm GC} &=& 17^{\rm h}45^{\rm m}40.045^{\rm s},\\
  \delta_{\rm GC} &=& -29^{\circ}0^{'} 28.1^{''}.
\end{eqnarray}
 The relations among $\alpha$ and  $\delta$ and the ecliptical longitude $\lambda$ and latitude $\beta$ are, from standard spherical trigonometry \cite{Roy05},
\begin{eqnarray}
  \sin\beta &=& \cos\epsilon\sin\delta-\sin\alpha\cos\delta\sin\epsilon = -0.097, \lb{uba}\\
  \cos\lambda\cos\beta &=& \cos\alpha\cos\delta = -0.054, \\
  \sin\lambda\cos\beta &=& \sin\epsilon\sin\delta +\sin\alpha\cos\delta\cos\epsilon = -0.99,\lb{truba}
\end{eqnarray}
where $\epsilon=23.43$ deg is the obliquity of the Earth's equator to the ecliptic. Thus, since, by definition, $0^{\circ}\leq \lambda \leq 360^{\circ}$  and $-90^{\circ}\leq\beta\leq +90^{\circ}$,
\rfr{uba}-\rfr{truba} yield for GC
\begin{eqnarray}
  \lambda_{\rm GC} &=& 266.744^{\circ}, \lb{lagc}\\
  \beta_{\rm GC} &=& -5.407^{\circ}\lb{latgc}
\end{eqnarray}
which tell us that GC is approximately directed in the opposite direction of the ICRF $x$ axis.
The GC longitude and latitude yield for the tidal parameter $\mathcal{K}$ of a hypothetical X/Nemesis object
\begin{eqnarray}
\mathcal{K}_{\rm Pit} &=& (2.1\pm 0.6)\times 10^{-26}\ {\rm s}^{-2},\lb{tid1}\\
\mathcal{K}_{\rm Fie} &=& (3.5\pm 2.8)\times 10^{-26}\ {\rm s}^{-2}\lb{tid2},
\end{eqnarray}
or, equivalently, for the MOND quadrupole parameter $-q$
\begin{eqnarray}
-q_{\rm Pit} &=& 0.34\pm 0.10,\lb{mil1}\\
-q_{\rm Fie} &=& 0.6\pm 0.4.\lb{mil2}
\end{eqnarray}
Note that Milgrom in Ref. \cite{Mil09} predicts $10^{-2}\leq-q\leq 0.3$ for the relevant range of values for the Galactic field at the Sun's location, and for a variety of interpolating functions; thus, \rfr{mil1}-\rfr{mil2} tell us that
\eqi-q\geq 0.2;\eqf the upper bound is less tight being
\eqi-q_{\rm max}=0.4-1.\eqf
It must be noted that Milgrom in Ref. \cite{Mil09} made certain simplifications in his calculations that should be taken into account when comparing his results with ours. Indeed, in addition to $\lambda_{\rm GC}=180^{\circ},\beta_{\rm GC}=0^{\circ}$, he assumed perfectly ecliptic orbits, i.e. with $I=0^{\circ}$, obtaining only radial and transverse components of the perturbing acceleration. Thus, his precession of the longitude of perihelion $\varpi$ reduces to that of the argument of perihelion $\omega$ because there is no precession of the node $\Omega$.  Milgrom  in Ref. \cite{Mil09} acknowledges that such an approximation is not valid for bodies like Pluto and Icarus showing high inclinations to the ecliptic.
Actually, the quadrupolar field of X/EFE does induce a secular precession on $\Omega$ as well, as we will see in Section \ref{nodo}.

Let us, now, reason in terms of a rock-ice planetary body.
By assuming for it a mass as large as that of Mars
we have for its distance
\begin{eqnarray}
  r_{\rm Pit} &=& 84\pm 9\ {\rm au}, \\
  r_{\rm Fie} &=& 71\pm 19\ {\rm au}.
\end{eqnarray}
An Earth-sized body would be at
\begin{eqnarray}
  r_{\rm Pit} &=& 178\pm 20\ {\rm au}, \\
  r_{\rm Fie} &=& 150\pm 40\ {\rm au},
\end{eqnarray}
while a gaseous giant like Jupiter would be at
\begin{eqnarray}
  r_{\rm Pit} &=& 1.218\pm 0.135\ {\rm kau}, \\
  r_{\rm Fie} &=& 1.027\pm 0.274\ {\rm kau}.
\end{eqnarray}
The distance of a brown dwarf with $M=80M_{\rm J}$
would be
\begin{eqnarray}
  r_{\rm Pit} &=& 5.246\pm 0.583\ {\rm kau}, \\
  r_{\rm Fie} &=& 4.425\pm 1.180\ {\rm kau},
\end{eqnarray}
while an object with the mass of the Sun would be at
\begin{eqnarray}
  r_{\rm Pit} &=& 12.366\pm 1.374\ {\rm kau}, \\
  r_{\rm Fie} &=& 10.430\pm 2.781\ {\rm kau},
\end{eqnarray}
Incidentally, it may be of some interest to compute the tidal parameter of Sgr A$^{\ast}$ itself to see if its action could be detected from its influence on the motion of the solar system's planets. By assuming for it \cite{Rei04} $M_{\rm X}=4\times 10^6 M_{\odot}$ and $r_{\rm X}=8.5$ kpc, we have
\eqi \mathcal{K}_{\rm Sgr A^{\ast}} = 3\times 10^{-35}\ {\rm s}^{-2},\eqf
which is  9 orders of magnitude smaller than the present-day level of accuracy in measuring $\mathcal{K}$.

Let us, now, abandon the direction of GC, and, consequently, the MOND scenario, and look at $\mathcal{K}$ as a function of the ecliptic longitude and latitude of X without assuming any a-priori limitations on them.
Let us, first, use the Pitjeva result of \rfr{pitperi}. It turns out that the maximum value of $\mathcal{K}$, and, consequently, the minimum value for $r_{\rm X}$, occurs for
\begin{eqnarray}
  \lambda_{\rm X} &=& 18.3^{\circ}, \\
  \beta_{\rm X} &=& -89.9^{\circ},
\end{eqnarray}
i.e. perpendicularly to the ecliptic;
\eqi \mathcal{K}_{\rm max}=(4\pm 1)\times 10^{-26}\ {\rm s}^{-2}.\eqf
In this case, the distances of X, for different values of its postulated mass, are
\begin{eqnarray}
  r_{\rm Mars} &=& 67\pm 7\ {\rm au}, \\
  r_{\rm Earth} &=& 141\pm 15\ {\rm au}, \\
  r_{\rm Jupiter} &=& 969\pm 107\ {\rm au}, \\
  r_{\rm brown\ dwarf} &=& 4.175\pm 0.463\ {\rm kau}, \\
  r_{\rm Sun} &=& 9.841\pm 1.093\ {\rm kau}.
  \end{eqnarray}
  The minimum for $\mathcal{K}$, corresponding to the maximum for $r_{\rm X}$, occurs for
  \begin{eqnarray}
  \lambda_{\rm X} &=& 182.8^{\circ}, \\
  \beta_{\rm X} &=& 1.7^{\circ},
\end{eqnarray}
i.e. basically in the ecliptic;
\eqi \mathcal{K}_{\rm min}=(2.0\pm 0.7)\times 10^{-26}\ {\rm s}^{-2}.\eqf The heliocentric distances for X are as follows
\begin{eqnarray}
  r_{\rm Mars} &=& 84\pm 9\ {\rm au}, \\
  r_{\rm Earth} &=& 178\pm 19\ {\rm au}, \\
  r_{\rm Jupiter} &=& 1.220\pm 0.135\ {\rm kau}, \\
  r_{\rm brown\ dwarf} &=& 5.261\pm 0.584\ {\rm kau}, \\
  r_{\rm Sun} &=& 12.400\pm 1.377\ {\rm kau}.
  \end{eqnarray}
Such results are quite similar to those obtained for GC.

In the case of the result by Fienga et al. \cite{Fie09} of \rfr{fieperi}, the minimum and the maximum of the tidal parameter occur at the same location as before. The maximum is
\eqi \mathcal{K}_{\rm max}=(7\pm 5)\times 10^{-26}\ {\rm s}^{-2}\eqf and yields
\begin{eqnarray}
  r_{\rm Mars} &=& 57\pm 15\ {\rm au}, \\
  r_{\rm Earth} &=& 120\pm 32\ {\rm au}, \\
  r_{\rm Jupiter} &=& 817\pm 218\ {\rm au}, \\
  r_{\rm brown\ dwarf} &=& 3.521\pm 0.939\ {\rm kau}, \\
  r_{\rm Sun} &=& 8.300\pm 2.789\ {\rm kau}.
  \end{eqnarray}
  The minimum is
 \eqi \mathcal{K}_{\rm min}=(3.5\pm 2.7)\times 10^{-26}\ {\rm s}^{-2}\eqf and the corresponding maximum distances are
\begin{eqnarray}
  r_{\rm Mars} &=& 72\pm 19\ {\rm au}, \\
  r_{\rm Earth} &=& 151\pm 40\ {\rm au}, \\
  r_{\rm Jupiter} &=& 1.030\pm 0.274\ {\rm kau}, \\
  r_{\rm brown\ dwarf} &=& 4.437\pm 1.183\ {\rm kau}, \\
  r_{\rm Sun} &=& 10.459\pm 2.789\ {\rm kau}.
  \end{eqnarray}
  \section{The action of a distant body and of MOND's External Field Effect on the node of a planet}\lb{nodo}
 Calculating the secular precession of the node $\Omega$ by means of the standard Gauss perturbative approach with \rfr{acce} for generic values of $I,\Omega,\omega$ and of $\lambda_{\rm X},\beta_{\rm X}$ yields  a non-vanishing effect. Also in this case, the exact formula is rather cumbersome; it is
 \eqi\left\langle\dot\Omega\right\rangle_{\rm X} = -\rp{3\mathcal{K}\csc I}{4n\sqrt{1-e^2}} \mathcal{H}(I,\beta_{\rm X},\lambda_{\rm X},\Omega)\mathcal{J}(I,\beta_{\rm X},\lambda_{\rm X},\Omega,\omega),\lb{nodazzo}\eqf
 with
 \eqi \mathcal{H} = \cos I\sin\beta_{\rm X} -\sin I\cos\beta_{\rm X}\sin(\lambda_{\rm X}-\Omega),\eqf
 and
 \eqi \mathcal{J} = (-2-3e^2+5e^2\cos 2\omega)\mathcal{G} -5e^2\sin 2\omega\cos\beta_{\rm X}\cos(\lambda_{\rm X}-\Omega),\eqf
 in which
 \eqi\mathcal{G}=\sin I\sin\beta_{\rm X} + \cos I\cos\beta_{\rm X}\sin(\lambda_{\rm X}-\Omega).\eqf
Note that, in general, it is not defined for $I\rightarrow 0^{\circ}$.
If and when also the corrections $\Delta\dot\Omega$ to the standard node precessions will be estimated, \rfr{nodazzo} could be used together $\left\langle\dot\varpi\right\rangle_{\rm X}$ and $\Delta\dot\varpi$ to constrain $\lambda_{\rm X}$ and $\beta_{\rm X}$  by taking their ratio and comparing it to the predicted one which would be independent of $\mathcal{K}$ itself, being a function of $\lambda_{\rm X},\beta_{\rm X}$ alone. In particular, it would be possible to check if $\lambda_{\rm GC},\beta_{\rm GC}$ satisfy the equation
\eqi\rp{\Delta\dot\Omega}{\Delta\dot\varpi}= \mathcal{Y}(\lambda_{\rm X},\beta_{\rm X})\lb{rapo},\eqf where
\eqi \mathcal{Y}(\lambda_{\rm X},\beta_{\rm X}) \equiv \rp{\left\langle\dot\Omega\right\rangle_{\rm X}}{\left\langle\dot\varpi\right\rangle_{\rm X}}\eqf
is the theoretically predicted ratio. This would also be a crucial test for the form of the MONDian EFE proposed by Milgrom in Ref. \cite{Mil09}. Indeed, if $\lambda_{\rm GC},\beta_{\rm GC}$ would not satisfy \rfr{rapo}, it should be rejected. The opposite case would not yet represent an unambiguous proof of MOND  because there would still be room for the action of an ordinary planetary-sized body; clearly, should observational efforts aimed to detect it be infructuous, MOND would receive a strong support.
\section{Discussion and conclusions}
We have shown that a putative  distant body X, not yet discovered, would induce non-vanishing secular precessions of the longitudes of the perihelion and  the node of a known planet P of the solar system. In particular, the resulting perihelion precession would be retrograde so that it would be able to explain the anomalous perihelion precession of Saturn recently determined  from an analysis including radio-technical data from Cassini. An investigation of the tidal parameter of X as a function of its ecliptic longitude and latitude showed that its maximum value occurs for X located perpendicularly to the ecliptic, while  its minimum occurs for X lying in the ecliptic. Accordingly, it has been possible to determine the present-day distance of X for different postulated values of its mass. Rock-ice planets as large as Mars and the Earth would be at about 80 au and 150 au, respectively, while a Jupiter-like gaseous giant would be at approximately 1 kau. A typical brown dwarf ($M=80 M_{\rm J}$) would be at about 5 kau, while Sun-sized body would be at approximately 10 kau. If it is difficult to believe that a main-sequence Sun-like star exists at just 10 kau from us, the distances obtained for terrestrial-type planets are substantially in agreement with theoretical predictions existing in literature about the existence of such bodies which would allow to explain certain features of the Edgeworth-Kuiper belt. Incidentally, let us note that our results rule out the possibility that the hypothesized Nemesis can be the Sun-like object X that may be responsible of the anomalous perihelion precessions of Saturn, also because, at approximately just 10 kau from us, its orbital period would amount to 1-10 Myr, contrary to the 26 Myr periodicity in extinction rates on the Earth over the last 250 Myr which motivated the Nemesis proposal. Moreover, our Sun-sized body X would not penetrate the Oort cloud which is believed to extend from  50 kau to 150 kau. The tidal parameter of Nemesis would be, instead, $2-4$ orders of magnitude smaller than the present-day level of accuracy in measuring it ($10^{-26}$ s$^{-2}$). On the other hand, if our X had a distance of about 88 kau, as predicted for Nemesis, our result for its tidal parameter would imply a mass of $300 M_{\odot}$.

For a particular position of X, i.e. along the  direction of the Galactic Center, our results hold also for the recently proposed form of the External Field Effect in the framework of MOND in the sense that it would be able to explain the perihelion precession of Saturn in such a way that it mimics the existence of a body in the direction of the center of the Milky Way. The associated parameter $q$ ranges from 0.2 to $0.4-1$, while the theoretical predictions for various choices of the interpolating function and various values of the Galactic field at the Sun's location are $10^{-2}\leq-q\leq 0.3$.

Anyway, further data analyses of enlarged radio-ranging datasets from Cassini by different teams of astronomers are required to confirm the  existence of the anomalous perihelion precession of Saturn as a real physical effect needing explanation.

Finally, let us note that a complementary approach to the problem consists of re-analyzing all the planetary data with modified dynamical models explicitly including also  a planet X and solving for a dedicated parameter accounting for it.


\begin{thebibliography}{99}
%
\bibitem{And09}
Anderson JD, Nieto MM. Astrometric Solar-System Anomalies. In: Proceedings of American Astronomical Society, IAU Symposium $\#$261. Relativity in Fundamental Astronomy: Dynamics, Reference Frames, and Data Analysis 27 April - 1 May 2009 Virginia Beach, VA, USA. Available from: http://arxiv.org/abs/0907.2469
%
\bibitem{And08}
Anderson JD, Campbell JK, Ekelund JE, Ellis J, Jordan JF.  Anomalous Orbital-Energy Changes Observed during Spacecraft Flybys of Earth. Phys Rev Lett 2008; 100: 091102.
%
\bibitem{And98}
Anderson JD, Laing PA, Lau EL, Liu AS, Nieto MM, Turyshev SG. Indication, from Pioneer 10/11, Galileo, and Ulysses Data, of an Apparent Anomalous, Weak, Long-Range Acceleration. Phys Rev Lett 1998; 81: 2858-61.
%
\bibitem{Kra04}
Krasinsky GA, Brumberg VA. Secular increase of astronomical unit from analysis of the major planet motions, and its interpretation. Celest Mech Dyn Astron 2004; 90: 267-88.
%
\bibitem{Wil08}
Williams JG, Boggs DH. Lunar Core and Mantle. What Does LLR See?. In: Schillak S, Ed. Proceedings of the 16th International Workshop on Laser Ranging 13-17 October 2008 - Pozna\'{n}, Poland.
%
\bibitem{Ior09}
Iorio L. The Recently Determined Anomalous Perihelion Precession of Saturn. Astron J 2009; 137: 3615-18.
%
\bibitem{Roy05}
Roy AE. Orbital Motion. Fourth Edition. Institute of Physics: Bristol 2005.
%
\bibitem{Pit09}
Pitjeva EV. EPM Ephemerides and Relativity. Paper presented at the American Astronomical Society, IAU Symposium $\#$261. Relativity in Fundamental Astronomy: Dynamics, Reference Frames, and Data Analysis 27 April - 1 May 2009 Virginia Beach, VA, USA.
%
\bibitem{INPOP}
 Fienga A, Laskar J, Morley T, {\it et al.} INPOP08, a 4-D planetary ephemeris: From asteroid and time-scale computations to ESA Mars Express and Venus Express contributions.
Astron Astrophys 2009; doi:10.1051/0004-6361/200911755.
%
\bibitem{Fie09}
Fienga A, Laskar J, Kuchynka P, {\it et al.} Gravity tests with INPOP planetary ephemerides. Submitted to proceedings of American Astronomical Society, IAU Symposium $\#$261. Relativity in Fundamental Astronomy: Dynamics, Reference Frames, and Data Analysis 27 April - 1 May 2009 Virginia Beach, VA, USA.  Available from: http://arxiv.org/abs/0906.3962.
%
\bibitem{Pit08}
Pitjeva EV.  Ephemerides EPM2008: the updated model, constants, data. Paper presented at Journ\'{e}es \virg{Syst\`{e}mes de r\'{e}f\'{e}rence spatio-temporels}
and X. Lohrmann-Kolloquium
22-24 September 2008 - Dresden, Germany.
%
\bibitem{VLBA}
Jones D, Fomalont E, Dhawan V, Romney J, Lanyi G, Border J. VLBA Observations of Cassini to
Improve the Saturn Ephemeris. Paper presented at URSI National Radio Science Meeting, Boulder, January 2009.
%
\bibitem{Lyk}
Lykawka PS., Mukai T. An Outer Planet Beyond Pluto and the Origin of the Trans-Neptunian Belt Architecture. Astron J 2008; 135: 1161-200.
%
\bibitem{Nem1}
Whitmire DP, Jackson AA. Are periodic mass extinctions driven by a distant solar companion? Nature 1984; 308: 713-5.
%
\bibitem{Nem2}
Davis M, Hut P, Muller A. Extinction of species by periodic comet showers. 1984,  Nature 1984; 308: 715-7.
%
\bibitem{Mul02}
Muller RA. Measurement of the lunar impact record for the past 3.5 b.y. and implications for the Nemesis theory.   Geol Soc of America  Special Papers 2002; 356: 659-65.
%
\bibitem{IorNem}
Iorio L. Constraints on planet X/Nemesis from Solar System's inner dynamics. Mon Not R Astron Soc 2009; doi:10.1111/j.1365-2966.2009.15458.x.
%
\bibitem{Mil83}
Milgrom M. A modification of the Newtonian dynamics as a possible alternative to the hidden mass hypothesis. Astrophys J 1983; 270: 365-70.
%
\bibitem{Mil09}
Milgrom M. MOND effects in the inner Solar system.  Mon Not R Astron Soc 2009; 399: 474-86.
%
\bibitem{Hog91}
Hogg D, Quinlan G, Tremaine S. Dynamical limits on dark mass in the outer solar system.  Astron J 1991; 101: 2274-86.
%
\bibitem{Ber03}
Bertotti B, Farinella P, Vokrouhlick´y D. Physics of
the Solar System. Kluwer: Dordrecht 2003.
%
\bibitem{Joh99}
Johnston KJ, de Vegt Chr. Reference Frames in Astronomy. Ann Rev Astron Astrophys 1999; 37: 97-125.
%
\bibitem{Bege}
Begeman KG.,  Broeils AH.,  Sanders RH. Extended rotation curves of spiral galaxies - Dark haloes and modified dynamics.  Mon Not R Astron Soc 1991; 249: 523-37.
%
\bibitem{Rei04}
Reid MJ, Brunthaler A. The Proper Motion of Sagittarius A$^{\ast}$. II. The Mass of Sagittarius A$^{\ast}$.  Astrophys J 2004; 616: 872-84.
\end{thebibliography}
\end{document}